\documentstyle [prb,aps,epsf]{revtex}
\begin{document}
\twocolumn[{
\widetext
\draft

\title{
Non-Fermi-liquid behavior due to short range order
}

\author{Ross H. McKenzie\cite{email} and  David Scarratt}

\address{School of Physics, University of New
South Wales, Sydney 2052, Australia}

\date{Received 23 May 1996}
\maketitle
\mediumtext
\begin{abstract}
An exactly soluble one-dimensional model of electrons
interacting with order parameter fluctations
associated with short-range order is considered.
The energy and momentum dependence of the electronic
self energy and spectral function are calculated
and found to exhibit non-Fermi-liquid features
similar to that seen for the two-dimensional
Hubbard model: a pseudogap, shadow bands,
anomalies in the self energy, and breakdown
of the quasiparticle picture.
Deviations from Fermi liquid behavior are      
largest close to the Fermi surface and  as the correlation length increases.
\\
\\
To appear in Physical Review B, Rapid Communications, November 1, 1996.
\end{abstract}
%74.20.Mn Nonconventional mechanisms (spin fluctuations, polarons, and
%         bipolarons, resonating valence bond model, anyon mechanism,
%         marginal Fermi liquid, Luttinger liquid, etc.)
%79.60.-i Photoemission and photoelectron spectra
%71.10.Fd Lattice fermion models (Hubbard model, etc.)
%71.10.Hf Non-Fermi-liquid ground states, electron phase diagrams and phase
%         transitions in model systems
%71.10.Li Excited states and pairing interactions in model systems
%71.10.Pm Fermions in reduced dimensions (anyons, composite fermions,
%         Luttinger liquid, etc.)
%75.30.Fv Spin-density waves
%71.45.Lr Charge-density-wave systems

}] \narrowtext

The question as to whether the metallic
properties of high-$T_c$ superconductors\cite{rev}
and   quasi-one-dimensional materials\cite{1d}
can be described
by the Fermi liquid picture that works so well in conventional
metals has recently received considerable
attention.   Many theoretical studies have been made of
strongly correlated electron models such as the Hubbard and $t-J$
models \cite{dag}.  The availability 
of high quality photoemission
data \cite{SD95,A95,wells,larosa}
has recently focussed attention  on the electron spectral
weight function $A(k,E)$ which is related to the probability of observing an
electron with momentum $k$ and energy $E$.

A consistent picture is gradually emerging from studies of the 
two-dimensional Hubbard
model using quantum Monte Carlo simulations\cite{whi,bulut,haas,preuss}
 and studies using the
fluctuation exchange (FLEX) approximation \cite{flex,langer,deisz,sch}.
  Some common features are
observed as the temperature is lowered, the hole doping is
decreased, or the Coulomb repulsion $U$
is increased.  It has been suggested that these changes
have a common origin \cite{preuss}: they correspond to an increase in $\xi$, the
correlation length associated with short range
antiferromagnetic order \cite{lro}. 

The common non-Fermi-liquid features observed are:
(1) As  $\xi$ increases a pseudogap develops,
i.e., there is a suppression of the density of states near the
Fermi energy \cite{whi,bulut,haas,preuss,langer,deisz}.
(2) As  $\xi$ increases peaks in the electron
spectral function become smaller and
broader\cite{deisz}.
% This effect increases as the Fermi surface is
% approached \cite{bulut,haas,preuss,langer,deisz}.
Near the Fermi surface a breakdown of the quasiparticle
picture may occur.
(3) On the Fermi surface the real part of the self energy
$\Sigma(k,E)$ can develop
a positive slope at the Fermi energy\cite{deisz}. 
% This means that the
%quasiparticle weight,
% $Z \equiv (1- {\partial \Sigma(k,E) \over \partial E} )^{-1}$,
% defined in Fermi liquid theory is meaningless.
(4) The 
magnitude of the imaginary part of the self energy
has a local
maximum at the Fermi energy\cite{deisz}.
(5) ``Shadow bands'' exist due to the incipient antiferromagnetic order
\cite{KS90},
i.e., in addition to the peak in the spectral function at the energy $E$
and momentum $k$
% defined by the pole of the one-electron Green function that occurs
%near $E=\epsilon_k$ 
there is a much smaller peak at $k + Q$    where   $ Q = (\pi,\pi)$
 is the wave vector associated with antiferromagnetic
order (and the nesting vector of the Fermi surface for half filling).
This means the spectral function $A(k,E)$ has peaks for $|k|>k_F$ but $E<E_F$.
It is estimated that such shadow bands are observable
when the correlation length is larger than a couple of lattice spacings
\cite{haas}.
The corresponding photoemission peaks
 have been observed in metallic
$Bi_2Sr_2CaCu_2O_{8+x}$ \cite{A95,larosa}
and insulating $Sr_2CuO_2Cl_2$ \cite{wells}.
(6) The self energy associated with the shadow bands is singular \cite{Chu95}.
As $\xi$ increases the imaginary part of the
self energy is much larger on the shadow Fermi surface
than on the regular Fermi surface\cite{sch}.

%The above results have mostly been
%obtained by computationally intensive treatments.
%It is also desirable to develop analytic approaches which capture the
%essential physics. To help promote this we present a simple
The purpose of this Rapid Communication is
to point out that there is
a simple
 one-dimensional model which also has many of the features listed above.
It is easy to study
because it has an {\it exact} analytic solution,
found by Sadovskii \cite{sad}.
This solution can be used to test different approximation
schemes and methods of analytic continuation
used on two-dimensional models.
% Although it obviously cannot describe all the physics
%of the two-dimensional Hubbard model
%our study does provide insight      
%into the more complicated two-dimensional case.
Hopefully, the results presented here will 
also provide physical insight into how short range
order produces non-Fermi-liquid behavior in two dimensions.
The model is directly relevant to the related issues for
quasi-one-dimensional materials \cite{mck2,mck1}.

The model consists of left and right moving electrons, with Fermi
velocity $v_F$, interacting with a static backscattering random
potential $\Delta(z)$, with Hamiltonian 
\begin{equation}
H = \int dz \Psi^\dagger  \bigg[ - iv_F \sigma_3
{\partial \over \partial z} + \Delta(z) \sigma_+ +
\Delta(z)^*
\sigma_-\bigg] \Psi
\label{hamel}
\end{equation}
where $\sigma_3$ and
$ \sigma_{\pm} \equiv {1 \over 2} (\sigma_1 \pm i \sigma_2)$ are Pauli matrices.
The upper and lower components of the spinor $ \Psi(z) $ are
left-moving, up-spin and right-moving, down-spin electrons, respectively.
The other  electrons are described by a similar Hamiltonian.
(The form of the Hamiltonian is motivated by
considering backward scattering in the Hubbard model).
  The random potential has zero mean and 
finite range Gaussian correlations given by:
\begin{equation}
%\langle \Delta(z)\rangle = 0 \ \ \ \ ; \ \ \ \ \
\langle \Delta(z)\Delta(z')^* \rangle = \psi^2
\exp(-|z-z'|/\xi).
\label{cor2}
\end{equation}
where $\psi$ is the rms fluctuation in the potential at a given point and
$\xi$ is the correlation length.
 $\psi$ defines an energy scale and a length scale
 $\xi_0 \equiv v_F/\psi$.
It will be seen below that the ratio of the correlation length 
to $\xi_0$ determines to what extent
the short-range order
causes deviation from Fermi liquid behavior.
For the commensurate case of a half-filled band
$\Delta(z)$ is real.                

%  One of us used this model recently
%to consider the effect of charge-density-wave \cite{mck2} and spin-density-wave
% \cite{mck1} fluctuations on the electronic properties
%of quasi-one-dimensional materials. 
%The spectral function and frequency dependent
%conductivity for this model has also been considered by
%Sadovskii and Timofeev\cite{sad2}.

An alternative interpretation of the Hamiltonian is that it
represents electrons interacting with spin
fluctuations, $S(x)\equiv\Delta(x)/\psi$, with susceptibility
\begin{equation}
\chi(\omega,q)= \delta(\omega) \sum_{\pm} { \xi^{-1} \over
 (q \pm 2k_F)^2 + \xi^{-2} }
\label{susc}
\end{equation}
and $\psi$ then defines the    strength of the coupling
of the electrons to the spin fluctuations.
In this interpretation we are assuming that the energy
scale associated with the spin fluctuations is so much
smaller than the electronic energy scale
that the spin fluctations can be treated as {\it static.}
Deisz, Hess, and Serene \cite {deisz} noted that 
for the two-dimensional Hubbard model at half filling
anomalies in the
self energy were always accompanied by a spin fluctuation 
T-matrix $T_{\sigma\sigma}(q,\omega_n)$ 
that is strongly peaked near the wave vector $q=Q$ and 
Matsubura frequency $\omega_n = 0$.
In this regime there is a natural mapping onto
the model considered here. 
At half-filling the Fermi surface is square and has
perfect nesting so is somewhat ``one-dimensional''.
The one-dimensional momentum then corresponds
to that along the $(\pi,\pi)$ direction
in the two dimensional model.
$\psi^2$ corresponds to the weight of the peak
in $T_{\sigma\sigma}(q,0)$
within $\xi^{-1}$ of $Q$ (compare  equation
(5) in Ref. \cite{deisz}).

Sadovskii considered the model defined by (\ref{hamel}) and (\ref{cor2})
 and found an {\it exact} solution by summing
all the diagrams generated by perturbation theory \cite{sad}.
%The electronic self energy is a continued fraction
%defined by the recursion relation:
%\begin{equation}
% \Sigma_l(k,E)={\psi^2 a(l)\over G_0((-1)^l(k + il {\rm sign} k/\xi) ,E)^{-1}
% -  \Sigma_{l +1}(k,E) }           
%\ \ \ \ \ l=1,2,3, \cdots
%\label{rec}
%\end{equation}
The electronic Greens function is a continued fraction
\begin{equation}
 G(k,E)=
{\strut{ 1
 \over\displaystyle{ E - a_0 -
{\strut{\psi^2c_1}\over\displaystyle{E-a_1 -
{\strut{\psi^2c_2}\over\displaystyle{E - a_2 - \cdots}
} } }}             }}
\label{contfrac}
\end{equation}
 where for right moving electrons 
 \begin{equation} a_l=(-1)^l v_F k +  {\rm sign}(k) \ {i \ l \over \xi}
 \label{al}
\end{equation}
%\begin{equation}
%b_l^2= \psi^2 c(l)
%\label{bl}
%\end{equation}
and $c_l=l$ for the commensurate case of a half-filled band and
for the incommensurate case $c_l=l/2$ for $l$ even and 
$c_l=(l+1)/2$ for $l$ odd.
As far as we are aware this is the only non-trivial electronic
model
for which there is an exact analytic solution
for the continued fraction representation of
a correlation function.
Since the Lanczos method of exact
diagonalization produces such
a continued fraction representation of spectral functions\cite{dag}
this model could provide insights into trends in
the continued fraction
coefficients and possible termination procedures \cite{vis}.
%The Greens function for right moving electrons is given by
%\begin{equation}
% G(k,E)={1\over G_0(k,E)^{-1} -  \Sigma_1(k,E) }           
%\label{gf}
%\end{equation}
In the limit $\xi \to \infty$ a perturbation expansion for
$G(k,E)$ is divergent but Borel summable \cite{mck2}.
%The spectral function is 
%$ A(k,E)=-{1\over\pi}{\rm Im}\ G(k,E+i\eta)$
%where $\eta \to 0^+$.
We evaluated the continued fraction (\ref{contfrac})
numerically using the modified Lentz's method \cite{press}.
All the results shown here are for the incommensurate case.
Qualitatively similar results are obtained for the commensurate case.

Generally we find that as the correlation length increases and the Fermi
surface is approached the properties of the self energy and
spectral function deviate significantly from the quasiparticle picture
of Fermi liquid theory.  Many of the anomalous features we see are
similar to those found for the two-dimensional Hubbard model at half-
filling treated in the fluctuation-exchange approximation \cite{deisz}.
  Figure 1
shows how the spectral function $A(k,E)$
broadens significantly as the electron momentum approaches the Fermi
momentum.  This happens even for correlation lengths of the order of
$\xi_0=v_F/\psi$.  This means that for strong coupling, i.e., 
$\psi$  of the order of
the band width, the correlation length
$\xi$ can be of the order of a lattice constant.
Shadow bands are present and become larger as $\xi$ increases and $|k|$
decreases.

Figure 2 shows how the quasiparticle picture breaks down as the
correlation length $\xi$
increases.  As   $\xi$  increases from   $0.2 \xi_0 $ to
$100\xi_0 $   the
spectral function on the Fermi surface
evolves from a single narrow peak to two broad
peaks which are the precursors of conduction and valence bands
associated with long range spin-density-wave 
order.  For similar reasons as the
correlation length increases a pseudogap develops in the total density
of states (Figure 2 inset). 
It should be pointed out that the two peak structure
is {\it not} seen in FLEX results \cite{deisz}
or quantum Monte carlo for weak coupling on large
lattices \cite{whi}. Whether this absence of the two peaks is a
real property of the 2D Hubbard model or a result
of the FLEX approximation or not being able to go
to low enough temperatures is not clear.           

The self-energy has anomalous features similar to those found for the 2D
Hubbard model at half-filling \cite{deisz}.
  At the Fermi energy the real part of the
self energy has a positive slope (Figure 3). 
The magnitude of the imaginary part of the self energy, which in a
quasiparticle picture is related to the quasiparticle
lifetime and  develops a
maximum at the Fermi energy (Figure 3).
In contrast, in a Fermi liquid ${\rm Re}
\Sigma$ has a negative slope and
$|{\rm Im} \Sigma|$ has a minimum (which goes to zero as the temperature
goes to zero) at the Fermi energy.
 The opposite behavior observed here means that 
although for our model
there is a pole in the spectral function at  $E=0$
it is not possible to define a quasiparticle solution there.
%quasiparticle weight defined in Fermi liquid theory by 
% $Z \equiv (1- {\partial \Sigma(k,E) \over \partial E} )^{-1}$
%  is meaningless.

The self-energy on the Fermi surface
$\Sigma(0,E)$ develops a singularity at the
Fermi energy $E=0$ as the correlation length increases.
Figure 4 shows that $\Sigma(k,E)$ develops 
similar singular behavior near $E \sim -kv_F$
which corresponds to the position of the shadow band.
In general, the magnitude of the imaginary part of the self energy
(which is related to the scattering rate)
is much larger and more singular near $E \sim -kv_F$ than $E \sim +kv_F$. 
The scattering rate for the shadow band increases
with $\xi$, as is observed for the 2D Hubbard model
as the doping or temperature is decreased\cite{sch}.
Only for very large correlation lengths does
the shadow band feature correspond
to a pole of the spectral function, i.e.,
an energy $E_k$ which satisfies
$E_k - kv_F -{\rm Re} \Sigma(k,E_k)=0$.
Consequently, the shadow band peak is not a replica of 
the regular quasiparticle peak.
These  results can be compared to those of Chubukov \cite{Chu95}
for a two-dimensional spin-fluctuation model.

% Note that even for short correlation
%lengths the scattering rate does not have the quadratic energy
%dependence characteristic of a Fermi liquid.  Why?
% The scattering is due to disorder

Figure 5 shows the spectral function calculated for this
model using second order perturbation theory.
This corresponds to termination of the continued fraction
at $l=1$.
Such a perturbative form for the Greens function has been used
in calculations concerning the role of order parameter
fluctuations in quasi-one-dimensional materials \cite{lra}.
However, the discrepancy with the exact results shown
in Figure 1 shows this is quite unreliable 
for $|E| < \psi$ when $|k-k_F| < \psi /v_F$
and $\xi > v_F/\psi$.
In particular perturbation theory greatly underestimates
the width of the spectral function.
The inset of Figure 5 shows how perturbation
theory gives unreliable results for the
total density of states if $\xi >  \xi_0$.

In conclusion,
the simple model considered here has many features similar
to those seen for the two-dimensional Hubbard model
and provides insight into how
short range order can produce  non-Fermi-liquid behavior.
 Of particular interest is the following.
(a) The ratio of the correlation length to
$\xi_0 = v_F/\psi$ determines deviations
from Fermi liquid behavior.
% where $\psi$ is a measure of the coupling of the electrons
%to the order parameter fluctuations.
(b) Non-Fermi-liquid behavior occurs even when
the correlation length is sufficiently short
that there is only a weak   pseudogap.
(c) The ``shadow band'' is associated with singularities in
the self energy and only corresponds to poles
in the spectral function for very large correlation
lengths.
(d) Perturbation theory gives unreliable results
in the non-Fermi-liquid regime.

This work was stimulated by discussions with J. R. Schrieffer.
We  thank H. Castella, J. Deisz, D. Hess, M. Steiner,
 and J. Voit for very helpful discussions.
This work was supported by the Australian Research Council.

{\it Note added in proof.} Recent angle-resolved photoemission
measurements [H. Ding {\it et al.}, Nature {\bf 382}, 51 (1996);
A. G. Loeser {\it et al.}, Science {\bf 273}, 325 (1996)]
have measured a pseudogap in the normal state of
$Bi_2Sr_2CaCu_2O_{8+x}$. The observed opening up
of the pseudogap with decreasing doping and temperature 
are qualitatively similar to the how the pseudogap
considered here opens up with increasing correlation length
(Fig. 2 inset).

\newpage

%\begin{figure}
%\caption{
%(see equation (\protect\ref{aa1})).
%\label{figpsgap}}
%\end{figure}

\begin{figure}
\centerline{\epsfxsize=7.0cm \epsfbox{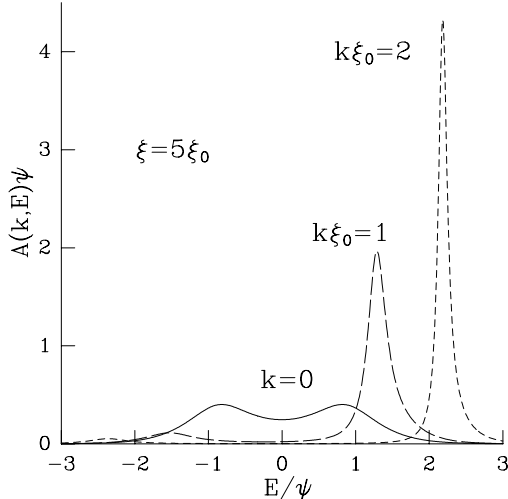}}
\caption{
Shadow bands and the breakdown of
 the quasiparticle picture near the Fermi surface.
The energy dependence of the electron spectral function $A(k,E)$ is shown
for three different momenta $k$.  The correlation length associated with
the short range order is equal to $5\xi_0=5v_F/\psi$  where 
$v_F $ is the Fermi velocity
and $\psi$  is a measure of the strength of the coupling of the electrons to
the order parameter fluctuations.  
Note the presence of ``shadow bands'', i.e.,
small peaks in $A(k,E)$ with $E < E_F=0$ and
$k> k_F=0$. The spectral function
becomes significantly broader as the Fermi surface is approached.
Opposite behavior occurs in a 
Fermi liquid:  the quasiparticles are better defined close to the Fermi
surface.  These results can be compared to those                     
for the two-dimensional Hubbard model at half filling
(see Figure 1 in Ref. \protect\cite{deisz}
and Figure 2 in Ref.      \protect\cite{haas}).
\label{fig1}}
\end{figure}

\begin{figure}
\centerline{\epsfxsize=7.0cm \epsfbox{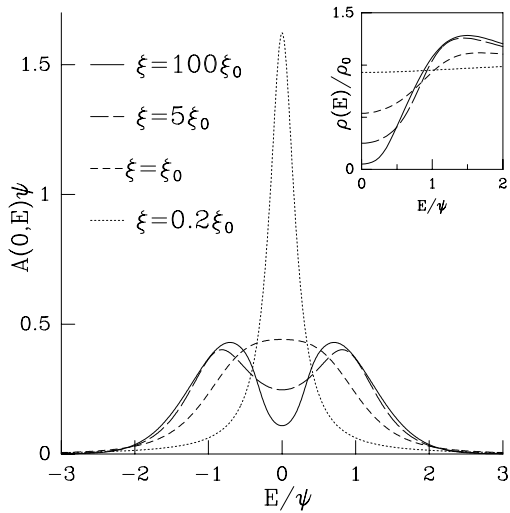}}
\caption{
Breakdown of the quasiparticle picture with increasing
correlation length.  The energy dependence of the spectral function at the
Fermi momentum is shown for several correlation lengths. The spectral
function broadens considerably and evolves into two bands as the
correlation length increases.  
These results can be compared to those                     
for the 2D
Hubbard model at half-filling as the  temperature is lowered (see
Figure 1 in Ref.      \protect\cite{deisz}).
The inset shows how the pseudogap in the total density of states opens
up as the correlation length increases.
Note that although for $\xi \sim \xi_0$ the
pseudogap is rather weak deviations from Fermi liquid behavior still
occur.
\label{fig2}}
\end{figure}

\begin{figure}
\centerline{\epsfxsize=7.0cm \epsfbox{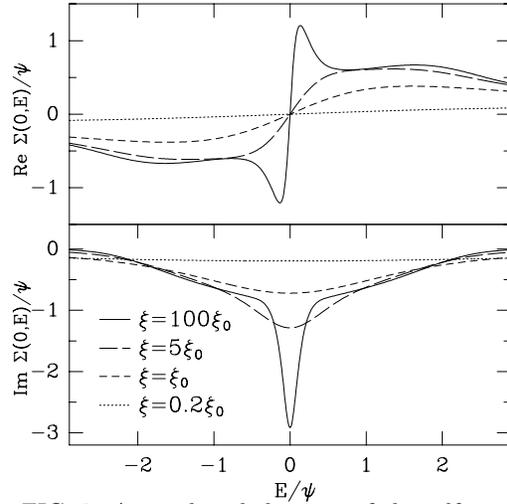}}
\caption{
Anomalous behaviour of the self energy.
The real part of the self energy at the Fermi momentum,
 ${\rm Re} \Sigma(0,E)$,
has a positive slope at the Fermi energy ($E=0$).
This means that a quasiparticle weight cannot be defined.
As the correlation length increases a maximum at $E=0$ develops
in the magnitude of the imaginary part of the self energy at the Fermi
momentum $|{\rm Im} \Sigma (0,E)|$. 
 (Compare Figure 2 in Ref.      \protect\cite{deisz}).
\label{fig3}}
 \end{figure}

\newpage
\begin{figure}
\centerline{\epsfxsize=7.0cm \epsfbox{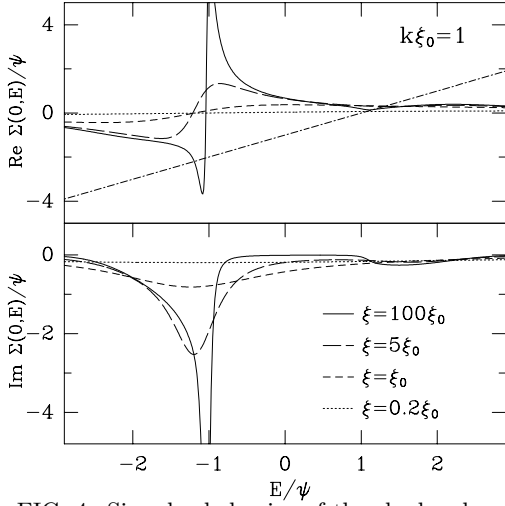}}
\caption{
Singular behavior of the shadow band self energy.
Away from the Fermi surface the self energy
near $E \sim k v_F$ 
has Fermi liquid properties.
In contrast, near $E \sim -k v_F$ which is associated
with the shadow band the self energy becomes
increasingly singular as the correlation length increases.
Intersections of the Re$\Sigma$ curves with the dot-dashed
straight line
correspond to poles of the spectral function.
Hence, the shadow band is only
associated with a pole for very large correlation lengths.
\label{fig4}}
 \end{figure}

\begin{figure}
\centerline{\epsfxsize=7.0cm \epsfbox{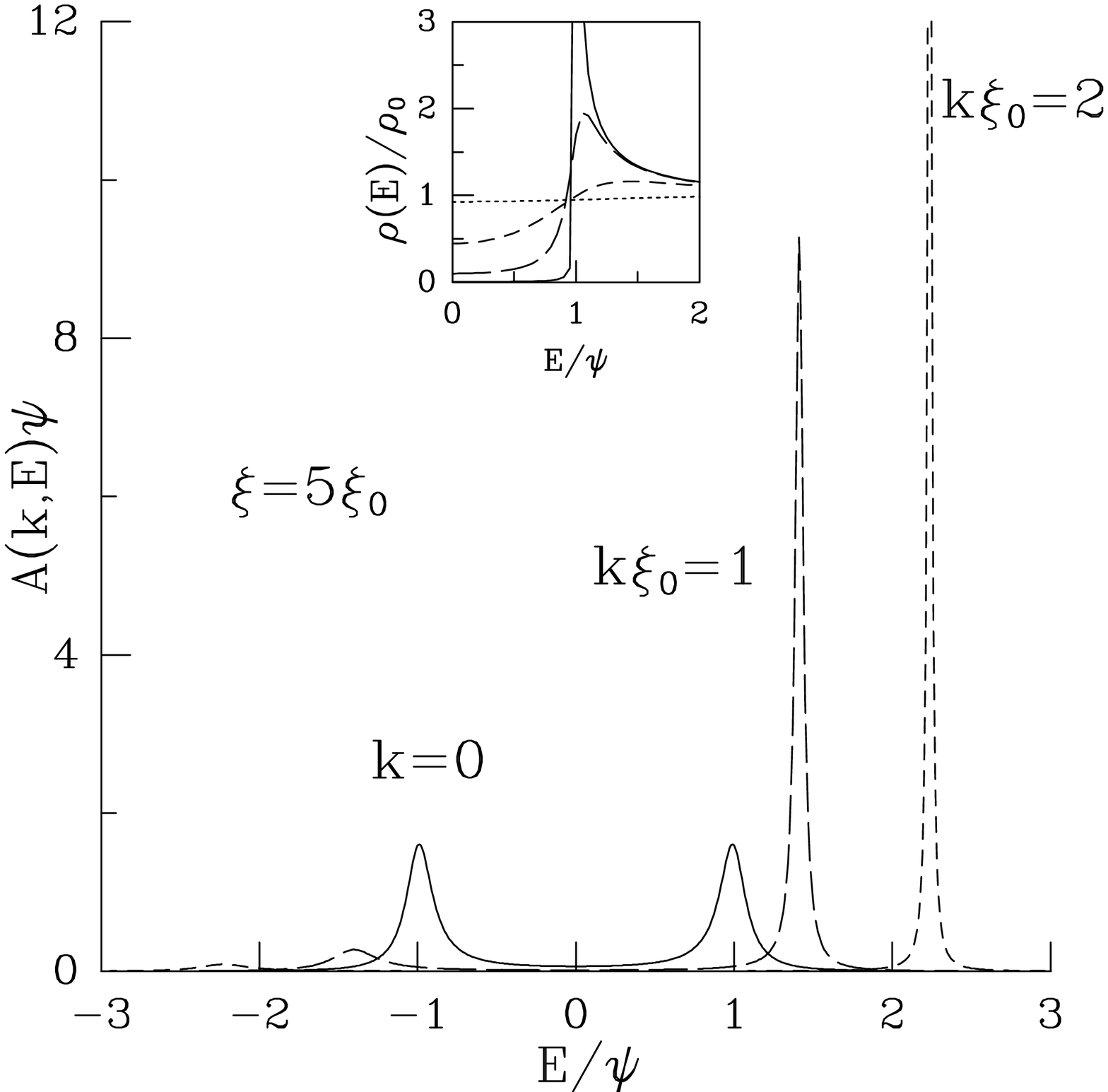}}
\caption{
Second order perturbation theory
is unreliable near the Fermi surface.
The spectral function is calculated for the same parameter values 
as in Figure 1.
 (Note the vertical scale is three times larger here).
 The discrepancy with the exact results
is large for $|k| < 1/\xi_0$ and becomes larger
as $\xi$ increases.
The inset shows the total density of states for
the same parameter values as in Figure 2.
 (Note the vertical scale is two   times larger here).
\label{fig5}}
\end{figure}


\begin{references}
\bibitem{lro}
 At zero temperature the ground state of the
Hubbard model at half-filling is believed to be a Mott-Hubbard
insulator with long-range antiferromagnetic order.  In two dimensions
this long range order is destroyed at finite temperature according
to the Mermin-Wagner theorem.
\end{references}
\end{document}